\documentstyle[12pt,epsfig]{article}
\def\lapprox{\lower .7ex\hbox{$\;\stackrel{\textstyle <}{\sim}\;$}}
\def\gapprox{\lower .7ex\hbox{$\;\stackrel{\textstyle >}{\sim}\;$}}

\def\d{{\rm d}}

\begin{document}
\begin{titlepage}
\vspace*{-1cm}
\begin{flushright}
DESY--99--077 \\
TTP99--29\\
June 1999 \\
\end{flushright}                                
\vskip 3.5cm
\begin{center}
{\Large\bf Azimuthal Asymmetries in}
\vskip 0.2cm
{\Large\bf Hadronic Final States at HERA}
\vskip 1.cm
{\large  M.~Ahmed$^{a,b}$} and {\large T.~Gehrmann$^c$} 
\vskip .4cm
{\it $^a$ II.~Institut f\"ur Theoretische Physik,
Universit\"at Hamburg, Luruper Chaussee 149, D-22761 Hamburg, Germany}
\vskip .2cm
{\it $^b$ University of the Punjab, Lahore, Pakistan}
\vskip .2cm 
{\it $^c$ Institut f\"ur Theoretische Teilchenphysik, Universit\"at 
Karlsruhe, D-76128 Karlsruhe, Germany}
\end{center}
\vskip 2.cm

\begin{abstract}
The distribution of 
hadrons produced in deeply inelastic electron--proton collisions 
depends on the azimuthal angle between lepton scattering plane and 
hadron production plane in the photon-proton centre-of-mass frame. 
In addition to the well known up-down 
asymmetry induced by the azimuthal dependence
of the Born level subprocess, there is also a non-vanishing left-right 
asymmetry, provided the incoming electron is polarized. This
asymmetry is time-reversal-odd and induced by absorptive corrections 
to the Born level process. We investigate the numerical magnitude 
of azimuthal 
asymmetries in semi-inclusive hadron production  
at HERA with particular emphasis on a possible determination of the 
time-reversal-odd  asymmetry. 
\end{abstract}

\vfill
\end{titlepage}                                                                
\newpage

The increase in the statistical accuracy of deep inelastic scattering (DIS)
data at the HERA collider will soon allow to investigate hadronic final state 
observables which go beyond hadron multiplicity 
distributions~\cite{haddistr} and jet rates
that have been studied up to now. Observables of particular interest are 
angular correlations between the lepton scattering plane, defined by 
the incoming and outgoing lepton momenta
 and the 
hadron production plane, defined by incoming proton and outgoing hadron
momentum. These correlations probe the dynamics and colour flow of the 
underlying partonic interaction at a detailed level, such that 
they can be used to test the perturbative description of hadron
production via partonic fragmentation.
An accurate prediction of the
perturbatively induced asymmetries is in particular desirable since
azimuthal correlations in semi-inclusive DIS have been suggested as
probes of non-perturbative effects in various places in the
literature~\cite{nplit}.

In this paper, we estimate the magnitude of the different azimuthal
asymmetries for kinematical
conditions at the HERA collider, using parton model 
expressions at leading order. Particular emphasis is put on
time-reversal-odd ($T$-odd) 
asymmetries, resulting from absorptive contributions
to the parton level scattering amplitudes~\cite{ru}. These $T$-odd
effects manifest in left-right asymmetries of the hadron
distribution with respect to the lepton scattering plane~\cite{men,hag}.
Corresponding to antisymmetric contributions to the hadronic tensor,
their observation requires either the contribution from parity violating 
weak interactions or polarization of the initial lepton beam. Given that 
lepton beam polarization will soon be realized for the HERA collider
experiments, both cases shall be investigated below. Perturbative
$T$-odd effects have up to now only been studied experimentally 
in polarized electron-positron
annihilation at SLAC~\cite{sld}, where the 
expected  asymmetries~\cite{bb} are however too
small to be measured directly, such that only upper limits
could be determined~\cite{sld}.

The kinematics of the semi-inclusive reaction
\begin{displaymath}
l(k) + p(p) \longrightarrow l'(k') + h(P) + X
\end{displaymath}
are  described by the following invariant variables
\begin{eqnarray}
Q^2 & = & -q^2 = -(k-k')^2\;,\nonumber \\
x &=& \frac{Q^2}{2q\cdot p}\;, \nonumber \\
z &=& \frac{p\cdot P}{q\cdot p}\;,\nonumber\\
\kappa^2 &=& z^2\left(1-\frac{q\cdot P}{xp\cdot P}\right) 
\end{eqnarray}
and the azimuthal angle $\phi$ between outgoing lepton direction and outgoing 
hadron direction measured in the centre-of-mass frame of virtual gauge
boson and proton. The variable $\kappa$ relates to the transverse
momentum of the outgoing hadron in this frame by $\kappa^2=P_T^2/Q^2$. 
The semi-inclusive scattering cross section can be decomposed according
to the dependence on $\phi$:
\begin{equation}
\frac{\d \sigma}{\d x \d Q^2 \d z \d \phi \d P_T^2} = 
\frac{\alpha^2 \pi}{2 Q^6 z} \left( A + B \cos \phi + C \cos 2\phi + D
  \sin \phi + E \sin 2\phi\right)\; .
\label{eq:master}
\end{equation}
Explicit parton model expressions for the coefficients $A$--$E$ can be
found in~\cite{men,hag}, the description of charged current (CC) interactions 
requires the substitution $\alpha \to G_F Q^2 /(\sqrt{2} \pi)$.
It should be noted that the leading order
contribution to $A$ is ${\cal O}(1)$, corresponding to vanishing
transverse momentum of the outgoing hadron. The first contribution to
$A$ yielding $\kappa \neq 0$ is ${\cal O}(\alpha_s)$.
The leading order contributions 
to $B$ and $C$ are ${\cal O}(\alpha_s)$ and to $D$ and $E$ are 
${\cal O}(\alpha_s^2)$. 

$D$ and $E$ are time-reversal-odd, they
are induced by absorptive one-loop corrections to the partonic  
scattering amplitudes. They appear in the hadronic tensor with 
asymmetric coefficients, which implies their vanishing for
purely electromagnetic interactions with unpolarized
beams. Non-vanishing $T$-odd asymmetries are obtained only for
weak interactions or for electromagnetic interactions with 
a longitudinally polarized lepton beam. 

In order to suppress the large ${\cal O}(1)$ contribution to the
$\phi$-independent coefficient $A$, it is appropriate to restrict
studies of angular asymmetries to hadrons produced at non-zero $p_T$. 
To project out individual terms in (\ref{eq:master}), we define the 
following average asymmetries, depending on $x$, $Q^2$ and $P_T$:
\begin{eqnarray}
\langle \sin (n\phi) \rangle (x,Q^2,P_T)
& = & \frac{\displaystyle \int \d z \d \phi \sin(n\phi)
\frac{\displaystyle \d \sigma}{\displaystyle \d x \d Q^2 \d z \d \phi
 \d P_T^2}}{\displaystyle
\int \d z \d \phi \frac{\displaystyle
\d \sigma}{\displaystyle \d x \d Q^2 \d z \d \phi
 \d P_T^2}}\; ,\nonumber \\
\langle \cos (n\phi) \rangle (x,Q^2,P_T)
& = & \frac{\displaystyle \int \d z \d \phi \cos(n\phi)
\frac{\displaystyle \d \sigma }{\displaystyle \d x \d Q^2 \d z \d \phi
 \d P_T^2}}{\displaystyle \int \d z \d \phi \frac{\displaystyle \d \sigma}
{\displaystyle \d x \d Q^2 \d z \d \phi
 \d P_T^2}}\; .
\label{eq:asydef}
\end{eqnarray}
The integration over the outgoing hadron momentum $z$ is a priori
bounded only by the kinematical requirement 
\begin{displaymath}
\kappa^2 \leq \frac{1-x}{x}z(1-z),
\end{displaymath}
which, at the $x$-values probed at HERA, involves contributions from
very small $z\ll 0.1$, where partonic fragmentation functions into
hadrons are only poorly ($0.01<z<0.1$) or not at all ($z<0.01$)
determined  from present experimental data. The uncertainty on the
fragmentation functions can however be expected to cancel to some extent
in the asymmetry, where fragmentation functions appear in numerator and
denominator. 
In our numerical studies below, we are dividing the
$z$-integration in (\ref{eq:asydef}) into five bins, in which we
compute the asymmetries.

The evaluation of the semi-inclusive DIS cross section (\ref{eq:master}) 
contains convolutions over parton distributions inside the target hadron 
as well as over fragmentation functions for outgoing partons into
observed charged hadrons. We use the
CTEQ4L~\cite{cteq}
leading order parton distribution functions, together with the
Binnewies-Kniehl-Kramer (BKK)~\cite{bkk}
 leading order fragmentation functions.
We restrict ourselves to charged hadron production (as also done in all 
experimental studies at HERA~\cite{haddistr} up to now)
and approximate the 
fragmentation function for a parton into a charged hadron by the sum
of the fragmentation functions to charged pions and kaons. This
approximation yields a satisfactory description of the ZEUS 
hadron production spectra in~\cite{haddistr}. 

Our evaluations for asymmetries in neutral and charged current (NC
and CC) exchanges assume $\sqrt{s}=300$~GeV and are made for several
points in the kinematical $(x,Q^2)$ plane accessible at this
centre-of-mass energy. 
For neutral current hadron production, we evaluate the angular
asymmetries in the bins listed in Table~\ref{tab:nc}. These bins all
correspond to $y=2/3$, where the $T$-odd contributions are kinematically 
largest. 
For the evaluation of $\langle \sin \phi\rangle$, we assume the electron 
beam to be left-handed; a right handed electron beam would result in an
asymmetry with opposite sign. 
\begin{table}[t]
\begin{center}
\begin{tabular}{|c||r|r|r|}\hline
Bin & 1 & 2 & 3 \\ \hline\hline
$\langle Q^2 \rangle/$GeV$^2$ & 6 & 60 &  600  \\ 
\hline
$\langle x\rangle$ & 0.0001 & 0.001 & 0.01  
\\ \hline
\end{tabular}
\end{center}
\caption{Kinematical bins used in the neutral current studies.}
\label{tab:nc}
\begin{center}
\begin{tabular}{|c||r|r|r|}\hline
Bin & 1 & 2 & 3  \\ \hline\hline
$\langle Q^2 \rangle/$GeV$^2$ & 1200 &  2000 & 5000  \\ \hline
$\langle x\rangle$ & 0.32 & 0.32 & 0.32 \\ \hline
\end{tabular}
\end{center}
\caption{Kinematical bins used in the charged current studies.}
\label{tab:cc}
\end{table}

\begin{figure}
\begin{center}
~ \epsfig{file=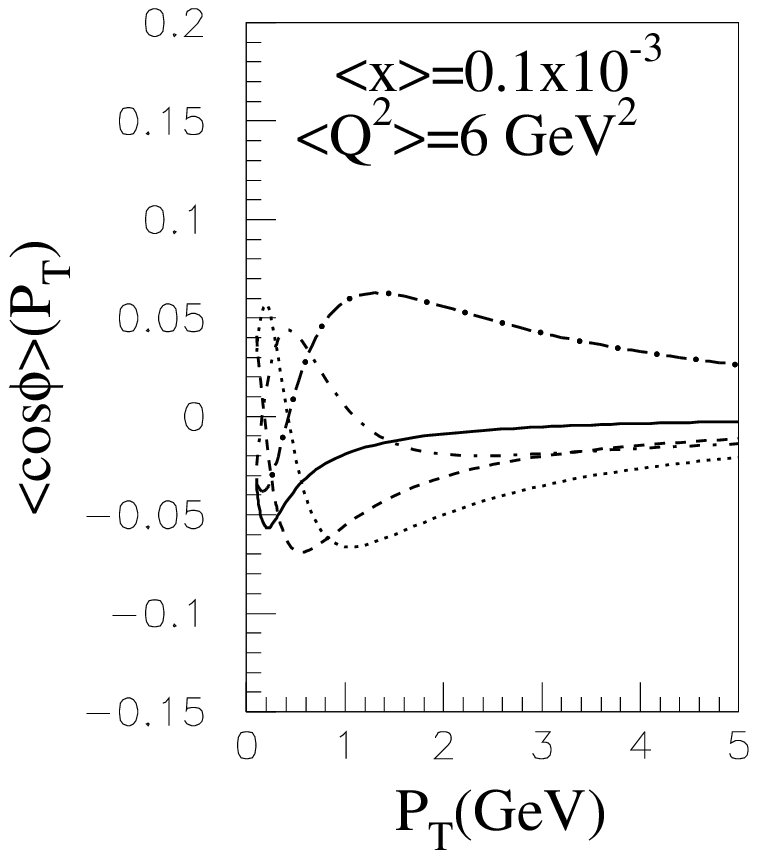,width=4cm}
~ \epsfig{file=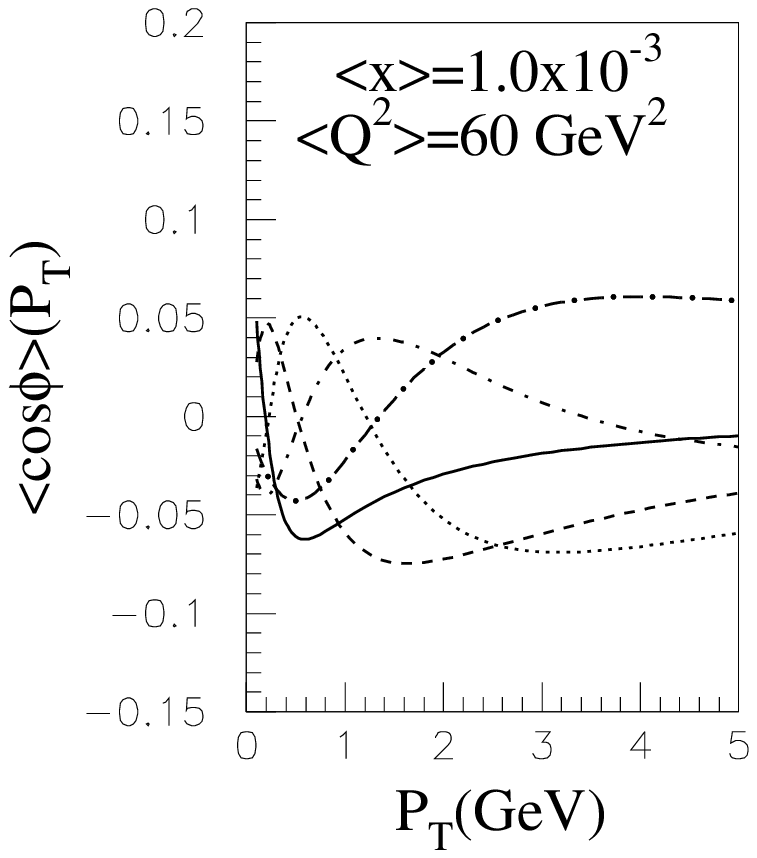,width=4cm}
~ \epsfig{file=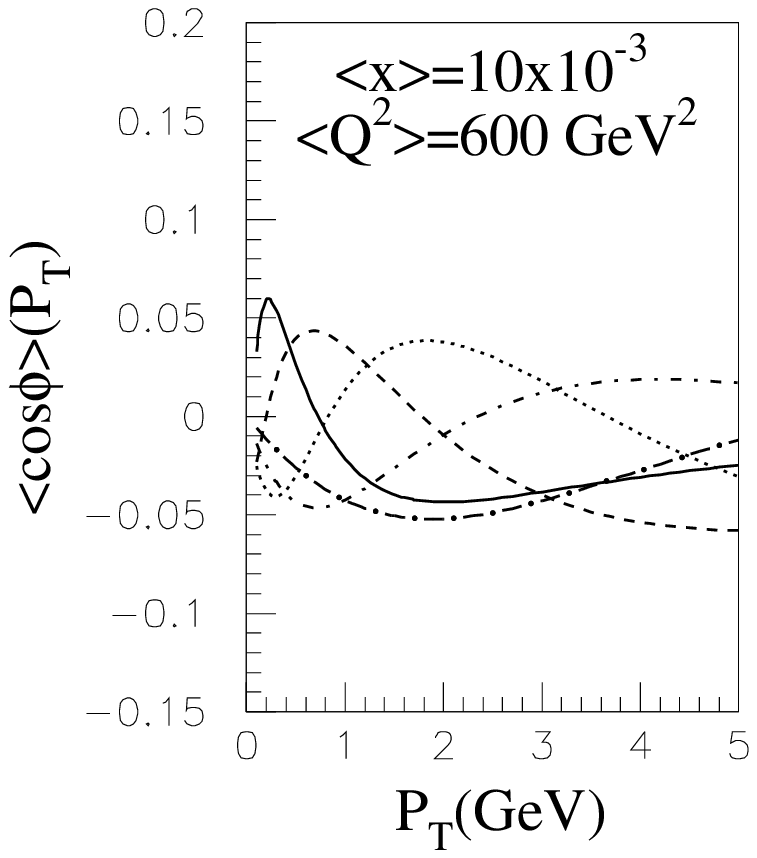,width=4cm}
\end{center}
\caption{The asymmetry \protect{$\langle \cos \phi\rangle(P_T)$} 
in neutral current  charged hadron production. 
Solid line: \protect{$0.005 < z < 0.01$}, 
dashed line: \protect{$0.01 < z < 0.05$}, 
dotted line: \protect{$0.05< z < 0.1$},
short dot-dashed line: \protect{$0.1<z<0.3$} 
and long dot-dashed line \protect{$0.3<z<0.9$}. }
\label{fig:nccos}
\end{figure}

\begin{figure}
\begin{center}
~ \epsfig{file=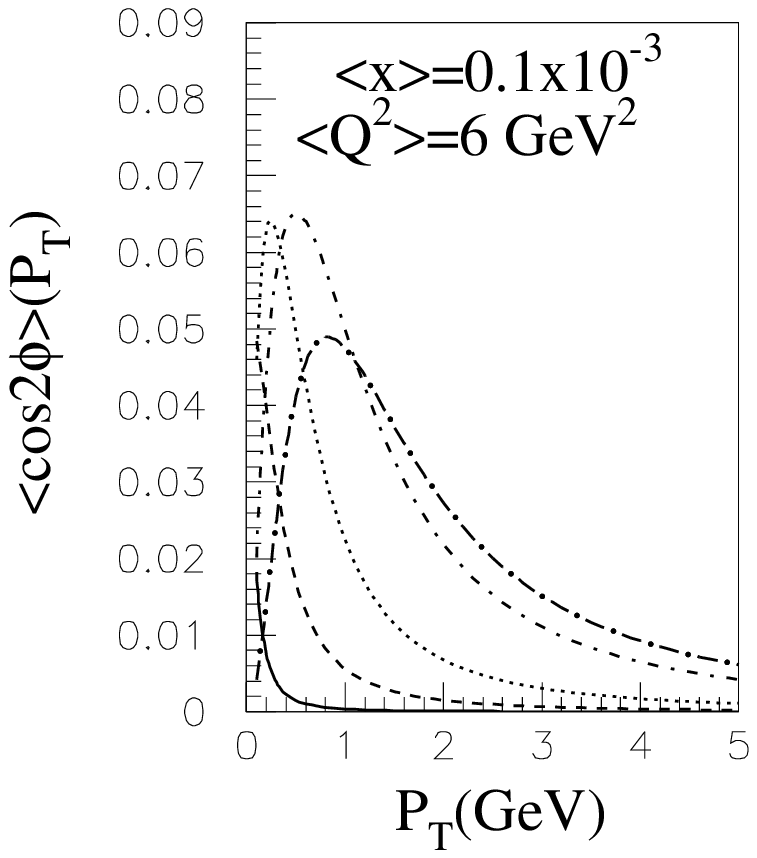,width=4cm}
~ \epsfig{file=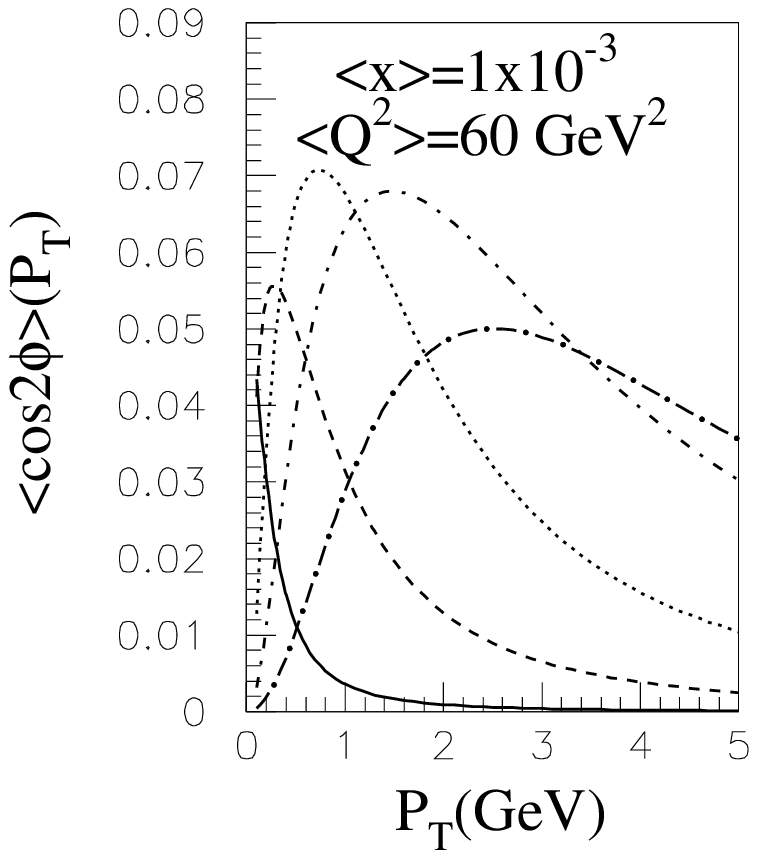,width=4cm}
~ \epsfig{file=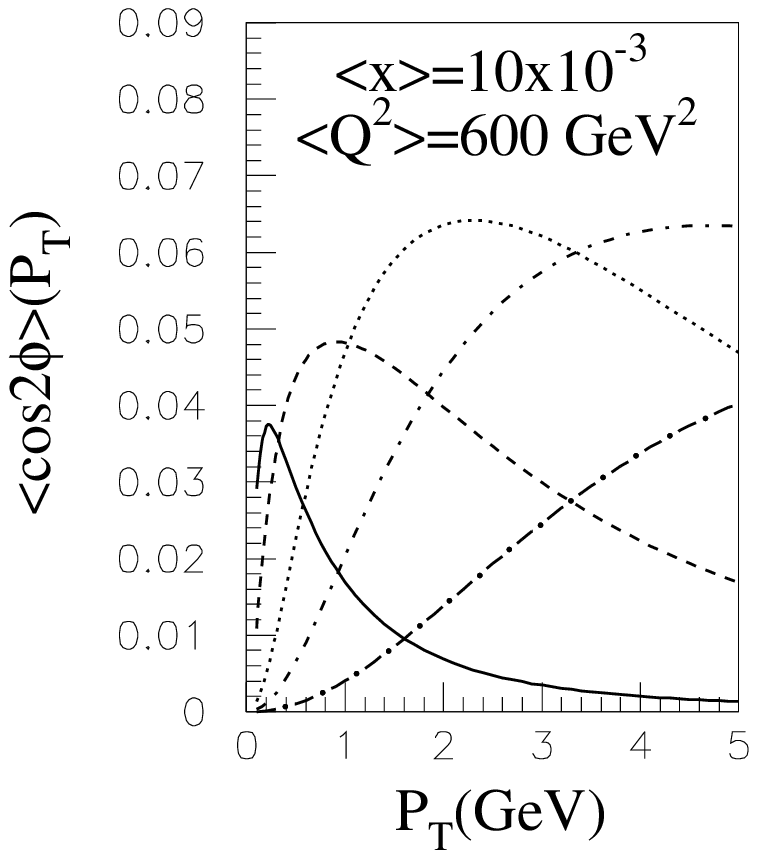,width=4cm}
\end{center}
\caption{The asymmetry 
\protect{$\langle \cos (2\phi)\rangle(P_T)$} in neutral current
  charged hadron production. Curves as in Fig.~\protect{\ref{fig:nccos}}.} 
\label{fig:nccos2}
\end{figure}

\begin{figure}
\begin{center}
~ \epsfig{file=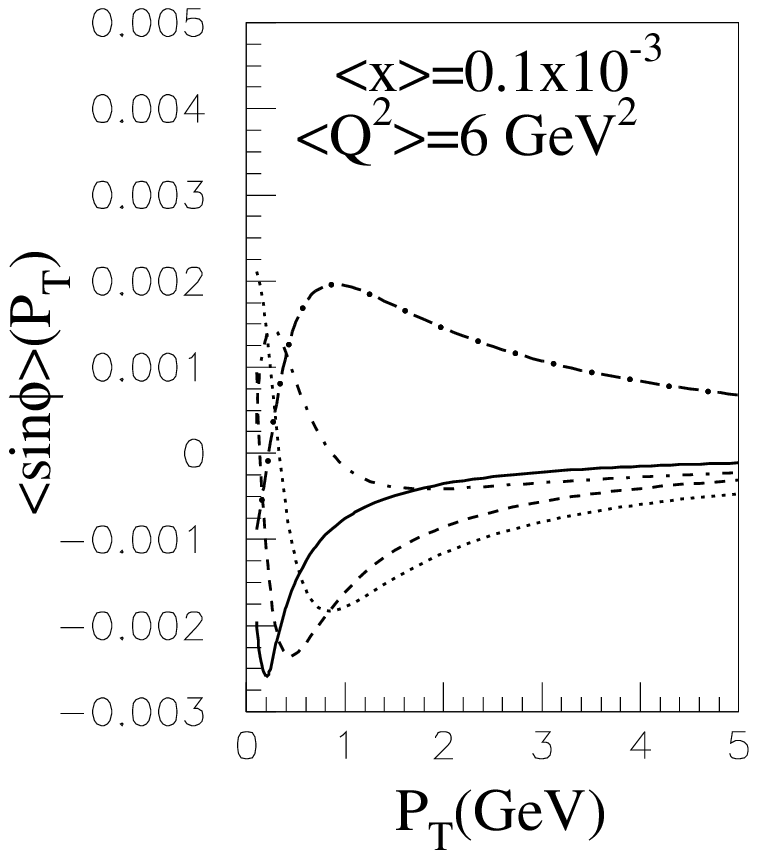,width=4cm}
~ \epsfig{file=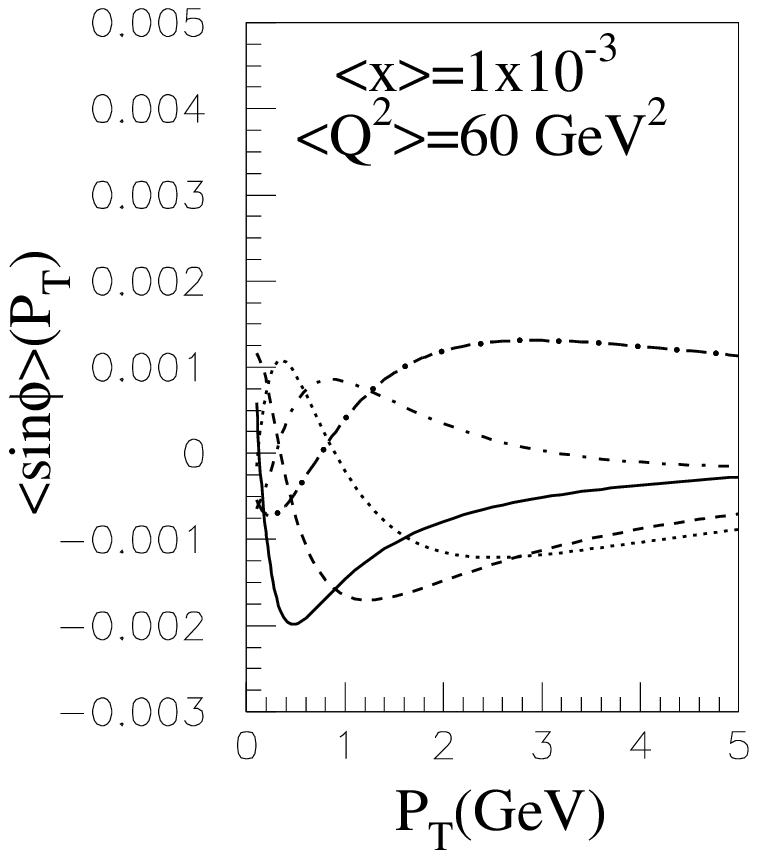,width=4cm}
~ \epsfig{file=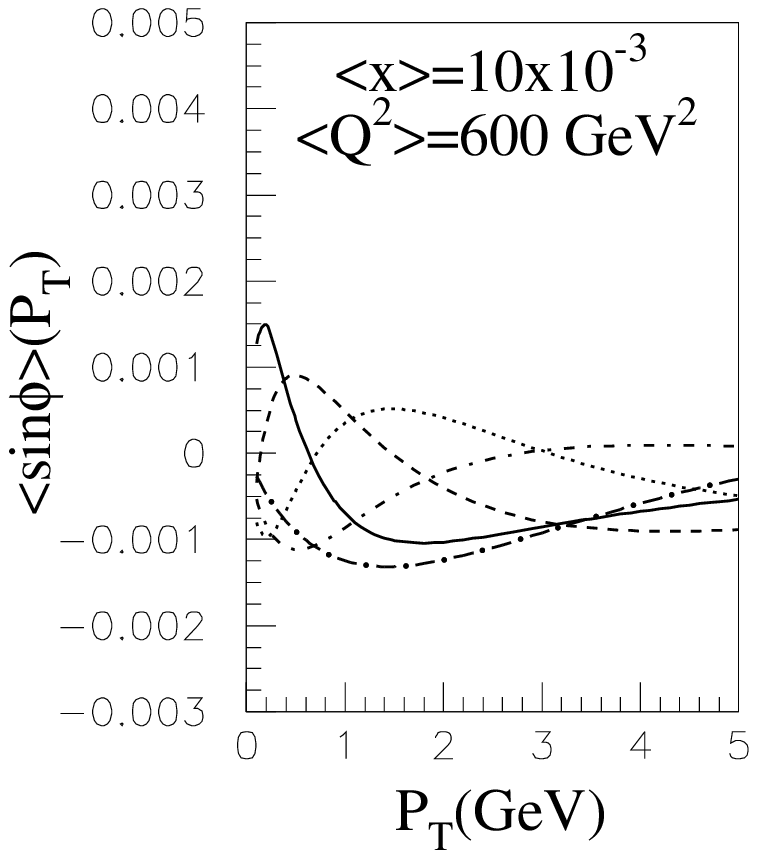,width=4cm}
\end{center}
\caption{The $T$-odd 
asymmetry 
\protect{$\langle \sin \phi\rangle(P_T)$} in neutral current
  charged hadron production. Curves as in Fig.~\protect{\ref{fig:nccos}}. }
\label{fig:ncsin}
\end{figure}

The resulting asymmetries are shown in
Figures~\ref{fig:nccos}--\ref{fig:ncsin}. It can be seen that $\langle
\cos \phi \rangle$ is typically of the order of a few per cent. For the
bins with lower $Q^2$, one observes that the asymmetry is positive 
and large for $0.3<z<0.9$, while almost vanishing for $0.1<z<0.3$. In
these $z$ bins, where the perturbative prediction is most reliable, 
this asymmetry should be easily accessible experimentally. $\langle
\cos (2\phi) \rangle$ appears to be of the same order of magnitude and
positive for all values of $z$. 

The $T$-odd asymmetry $\langle \sin \phi \rangle$ does not exceed two per
mille in the neutral current case, and attains its  largest values at
small $z$, where the fragmentation functions are only poorly
known. Given that this asymmetry has to be measured in the presence of 
$\langle \cos (n\phi) \rangle$-asymmetries, being about an order of 
magnitude larger,  it can only be concluded that the
determination of $T$-odd effects in semi-inclusive DIS is
experimentally challenging and requires large luminosity as well as good 
control over possible systematic effects linking 
$\langle \cos (n\phi) \rangle$ and $\langle \sin \phi \rangle$.
The estimated magnitude of two per mille should also be taken as a
reference value to be compared with non-perturbative
estimates~\cite{nplit}. 

The leading order 
expressions for semi-inclusive hadron production are easily generalized 
to jet production by replacing the fragmentation functions by a jet
definition. At lowest order, each parton can be identified with an
observed jet, such that a mere replacement of the fragmentation
functions by $\delta$-functions in the energy transfer yields
expressions for the 2+1 jet production cross section. Both jets are
produced back-to-back and with identical $P_T$ in the centre-of-mass
frame of gauge boson and proton, $\langle \cos (n\phi) \rangle$ and 
$\langle \sin (n\phi) \rangle$ vanish consequently in the integrated jet 
rate. Only by restricting
the final state configuration by cuts on the jet direction, such as
suggested for example in~\cite{kramer}, we obtain non-vanishing
asymmetries, which are comparable in magnitude to the asymmetries
obtained for hadron production. 

Charged current interactions at HERA result in a final state with an
undetected neutrino. Direction and energy of the neutrino can be 
inferred from the imbalance of momentum in the event, the reconstruction 
of the kinematical variables is however less precise than in the neutral 
current case. Charged current interactions are mediated by a
massive gauge boson, their magnitude becomes comparable to neutral
current interactions only towards large $Q^2$. Measurements of charged
current DIS at HERA can therefore be made only at large $Q^2$, and the
points for our numerical studies have been chosen accordingly. They
correspond to bin centres used in recent HERA measurements of the CC cross
section, and they are listed in Table~\ref{tab:cc}. In CC interactions,
it turns out that the $T$-odd asymmetries $\langle \sin (n\phi) \rangle$
become kinematically
largest for small $y$, such that we have selected bins
corresponding to a minimal value of $y$.

\begin{figure}
\begin{center}
~ \epsfig{file=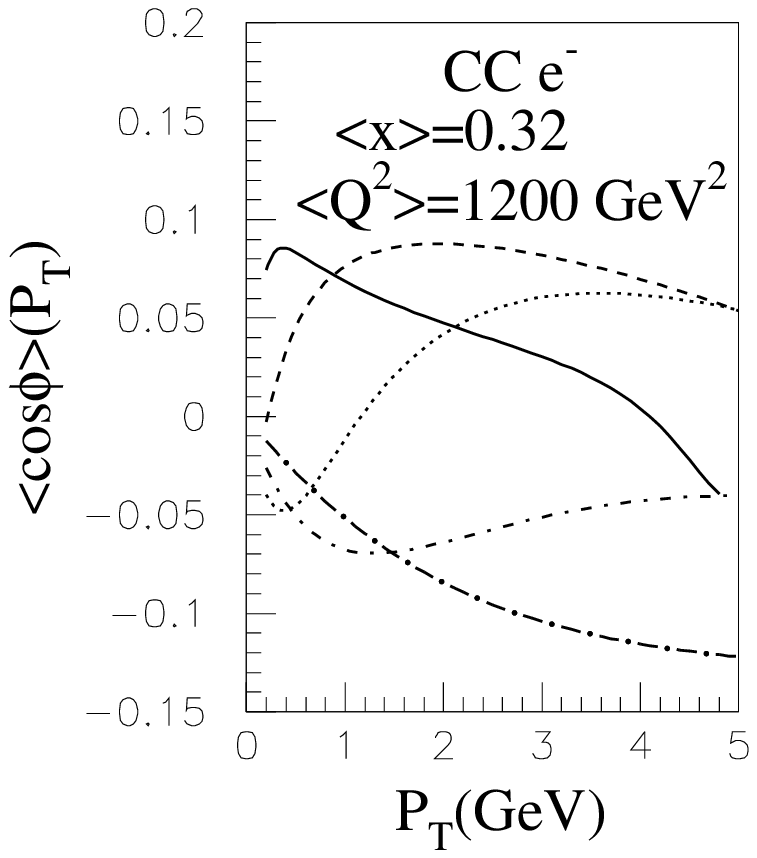,width=4cm}
~ \epsfig{file=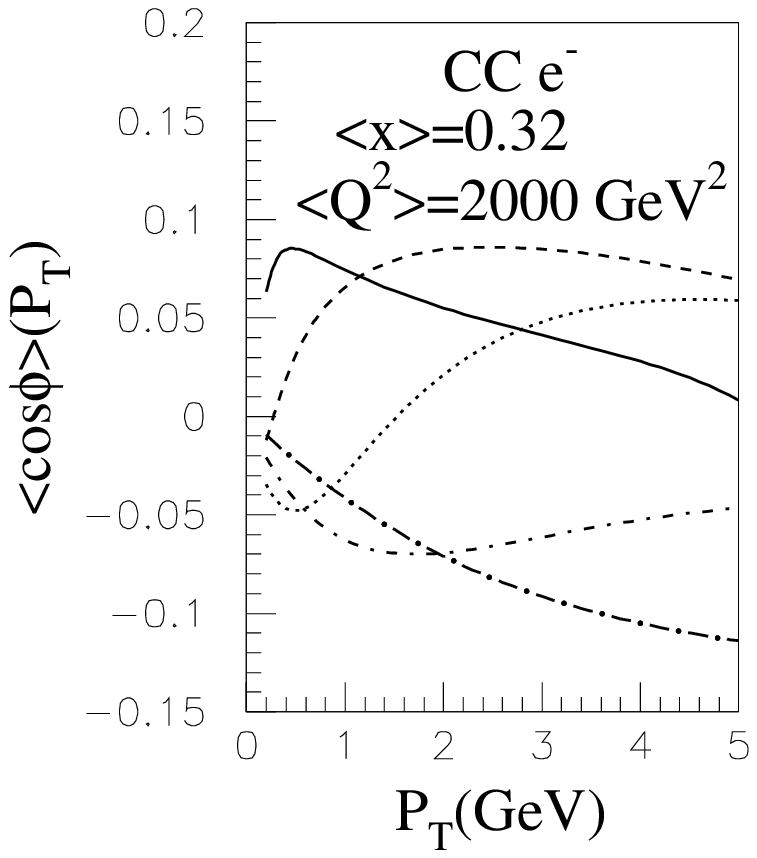,width=4cm}
~ \epsfig{file=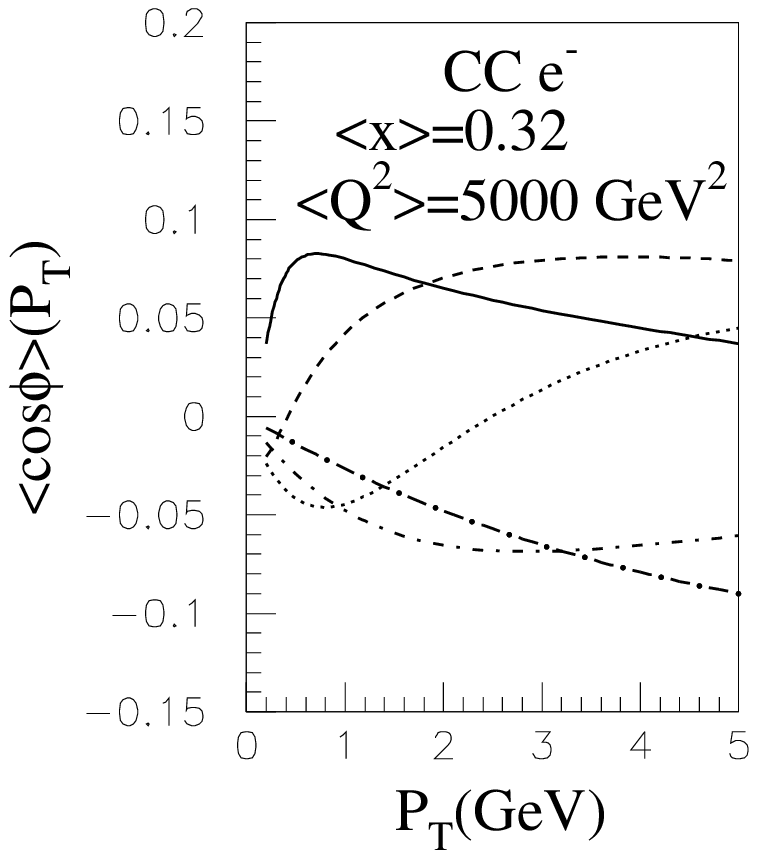,width=4cm}
\end{center}
\caption{The asymmetry \protect{$\langle \cos \phi\rangle(P_T)$} in charged
  current $(e^-)$
  charged hadron production. Curves as in Fig.~\protect{\ref{fig:nccos}}.} 
\label{fig:mcccos}
\end{figure}

\begin{figure}
\begin{center}
~ \epsfig{file=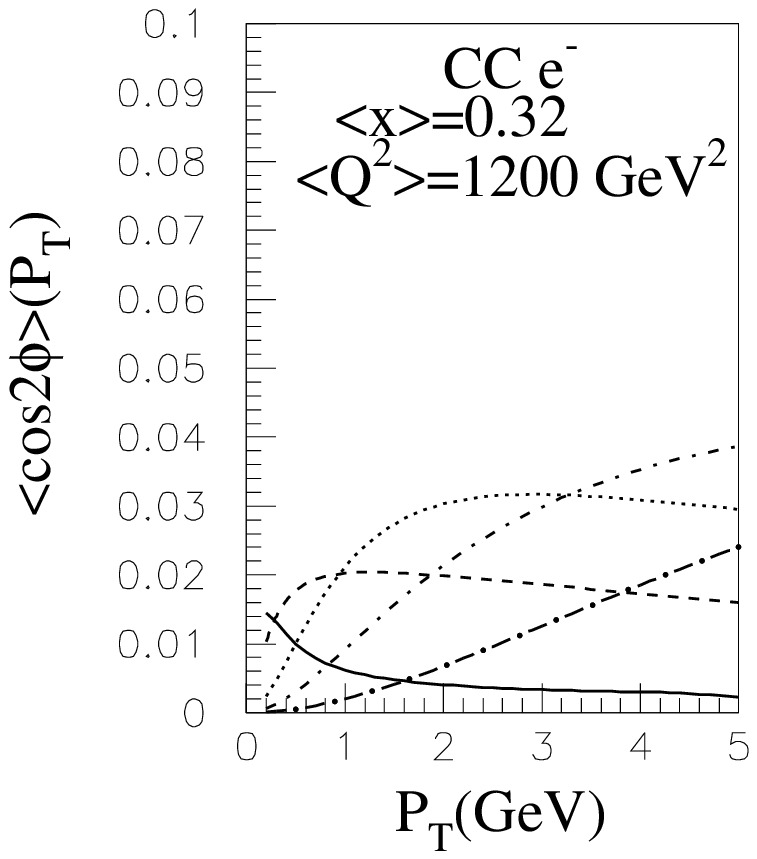,width=4cm}
~ \epsfig{file=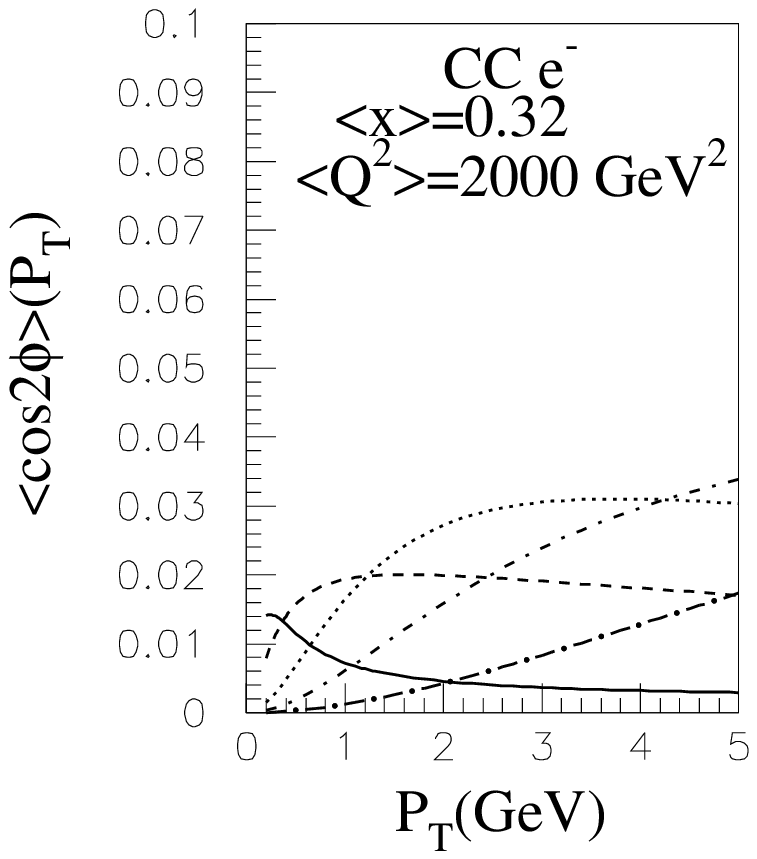,width=4cm}
~ \epsfig{file=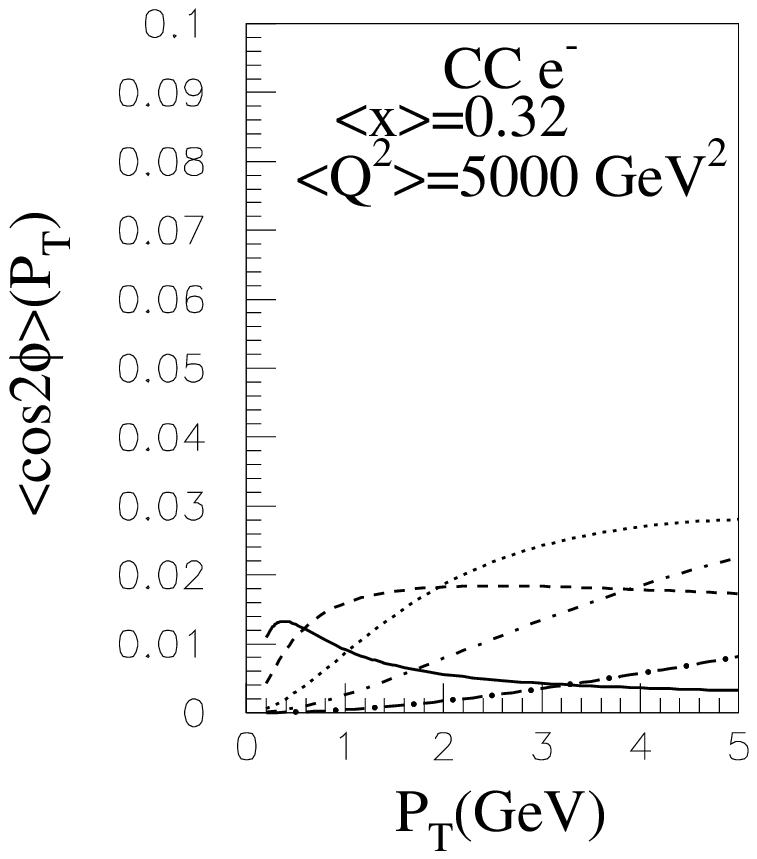,width=4cm}
\end{center}
\caption{The asymmetry \protect{$\langle \cos (2\phi)\rangle(P_T)$} 
in charged current
  $(e^-)$
 charged hadron production. Curves as in Fig.~\protect{\ref{fig:nccos}}.} 
\label{fig:mcccos2}
\end{figure}

\begin{figure}
\begin{center}
~ \epsfig{file=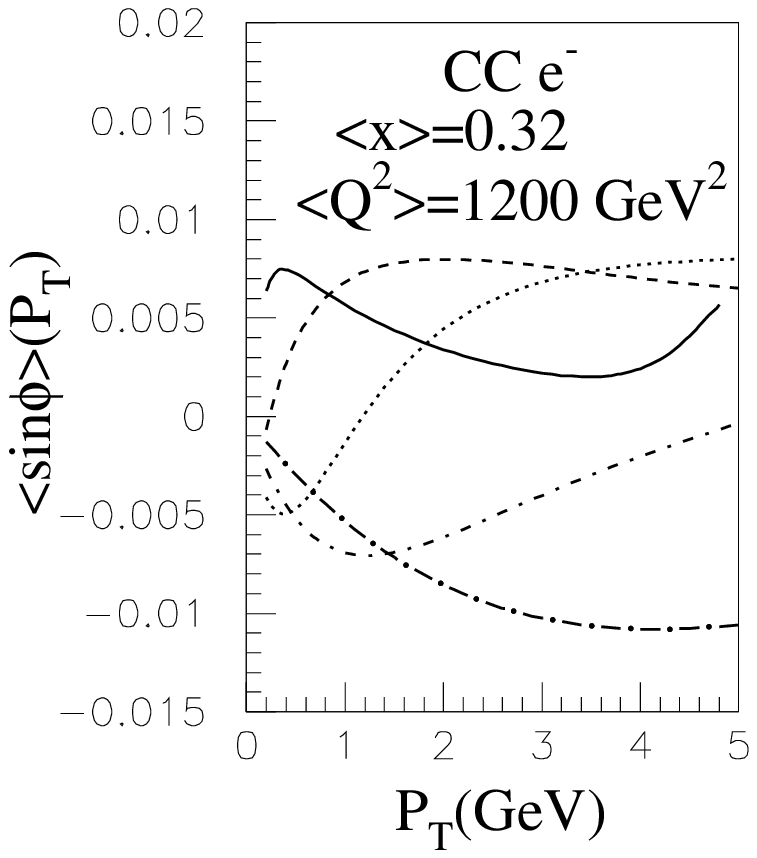,width=4cm}
~ \epsfig{file=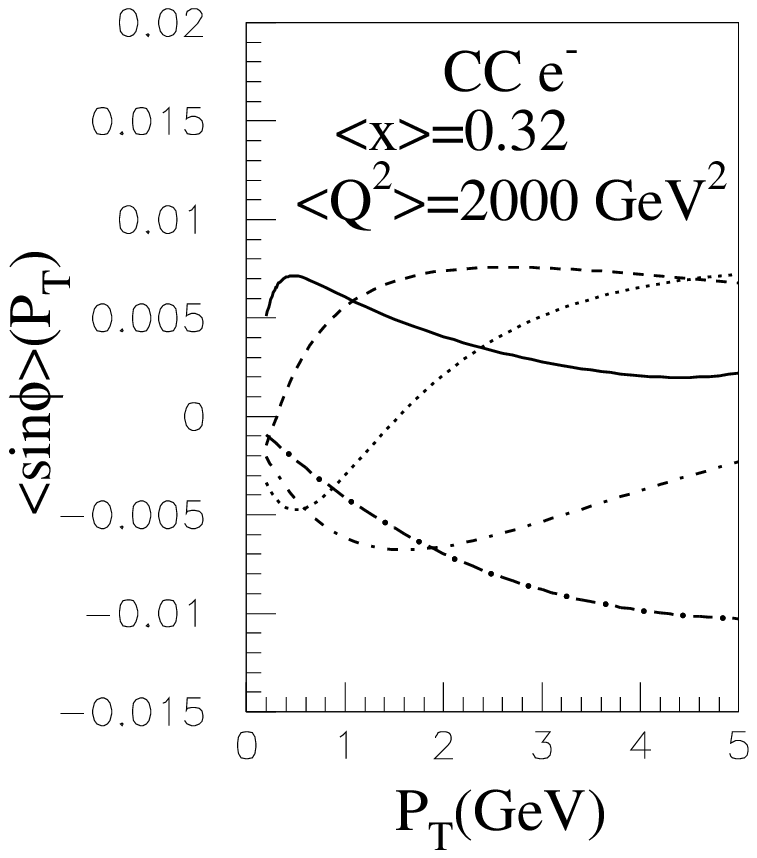,width=4cm}
~ \epsfig{file=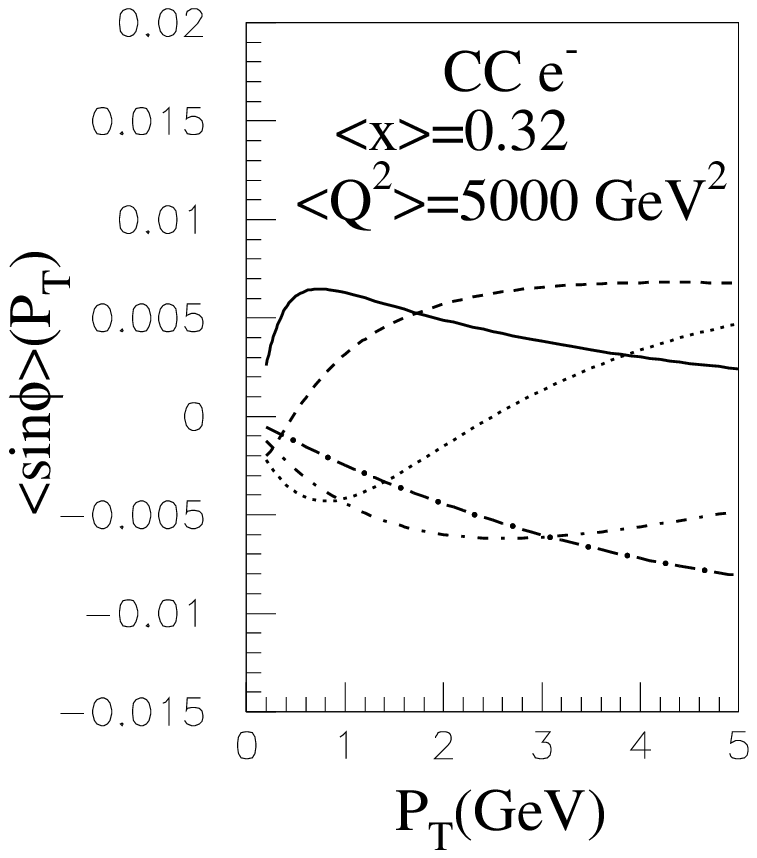,width=4cm}
\end{center}
\caption{The $T$-odd 
  asymmetry \protect{$\langle \sin \phi\rangle(P_T)$}
 in charged current  $(e^-)$
  charged 
hadron production. Curves as in Fig.~\protect{\ref{fig:nccos}}. }
\label{fig:mccsin}
\end{figure}

\begin{figure}
\begin{center}
~ \epsfig{file=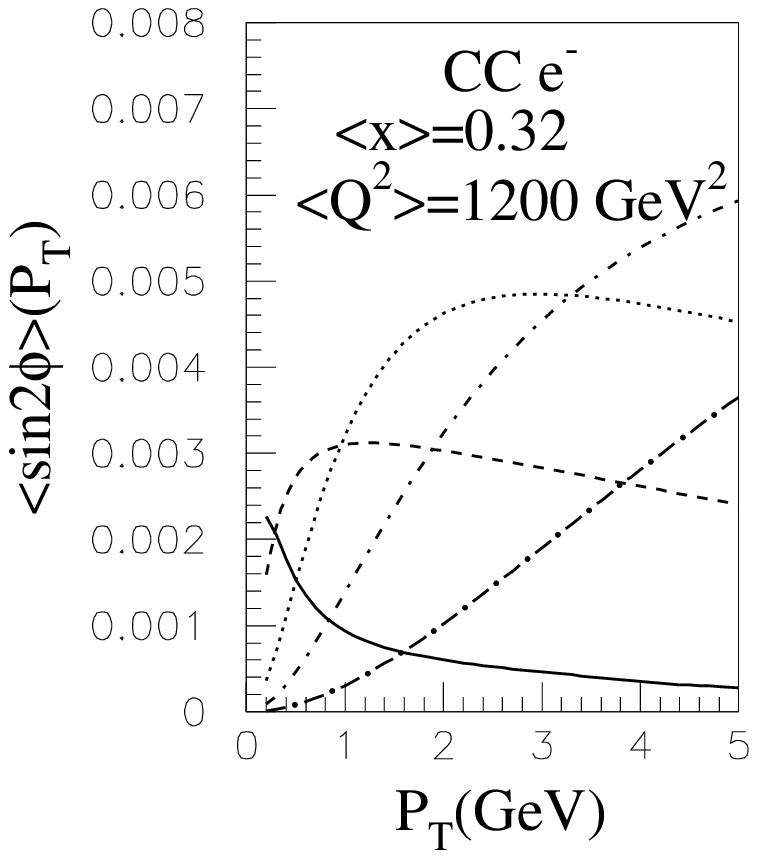,width=4cm}
~ \epsfig{file=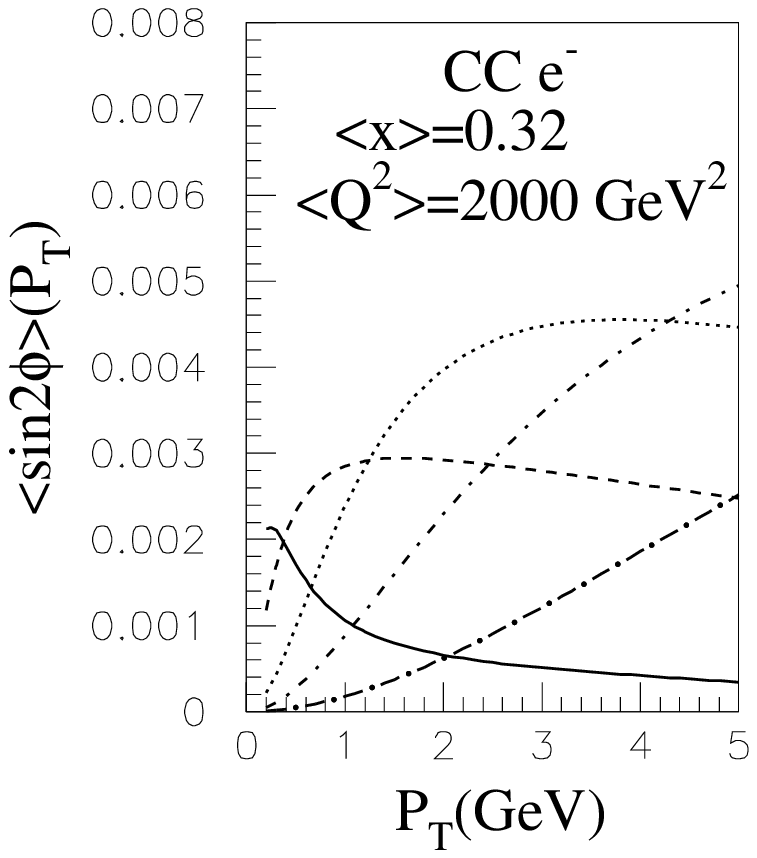,width=4cm}
~ \epsfig{file=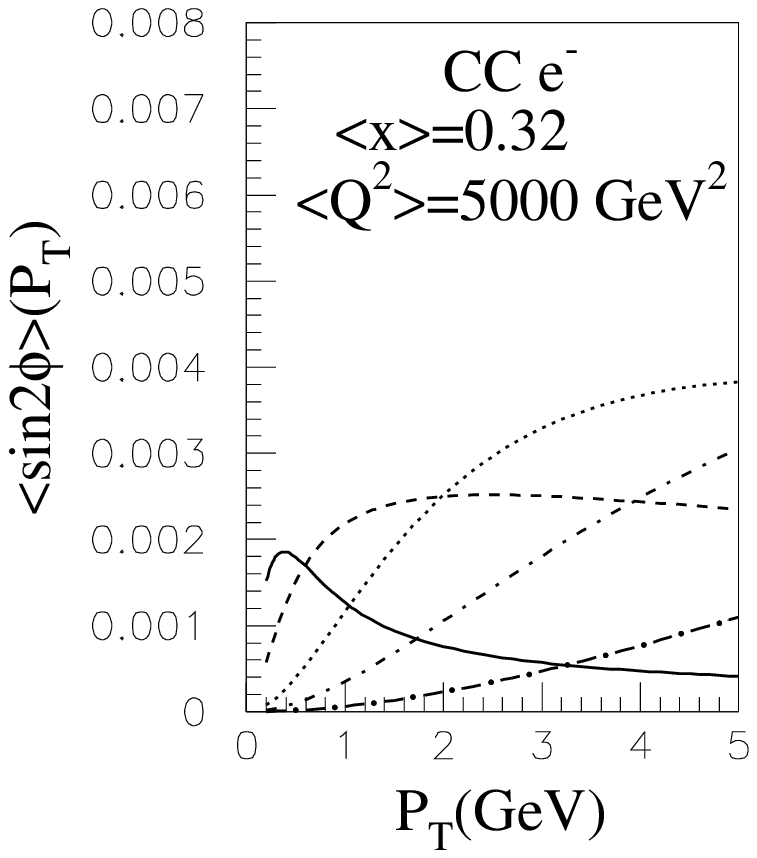,width=4cm}
\end{center}
\caption{The $T$-odd
asymmetry 
\protect{$\langle \sin (2 \phi)\rangle(P_T)$} in charged current $(e^-)$
  charged hadron production. Curves as in Fig.~\protect{\ref{fig:nccos}}.} 
\label{fig:mccsin2}
\end{figure}

For CC DIS, one obtains different cross sections for incoming positrons
and electrons, since the resulting $W^{\pm}$ currents 
couple to different combinations of quark distributions in the target. 
The cross section for electron scattering is larger than for positron
scattering, such that we shall only report on results 
for asymmetries in electron scattering here; the corresponding
asymmetries in positron scattering are smaller in magnitude due 
to the different sign of parity violating contributions, as
demonstrated in~\cite{hag}.
Figures~\ref{fig:mcccos}--\ref{fig:mccsin2} represent the azimuthal
asymmetries obtained for electron scattering. We observe a pattern
similar to the neutral current case: the $\langle\cos (n\phi)\rangle$
asymmetries are both of the order of several per cent, reaching maximum
values of about $10\%$ for
$\langle\cos (\phi)\rangle$ and $5\%$ for  $\langle\cos
(2\phi)\rangle$. 
In charged current DIS, two $T$-odd asymmetries
are present: $\langle\sin \phi\rangle$ and $\langle\sin (2\phi)\rangle$.
Both asymmetries are sizable for $z>0.1$, where 
the predictions are most reliable. $\langle\sin \phi\rangle$
turns out to be larger than in the neutral current case, and amounts
up to one per cent.  $\langle\sin (2\phi)\rangle$ is at the level
of half a per cent. 
The ratio between $\langle\cos (n\phi)\rangle$ and
$\langle\sin (n\phi)\rangle$ is therefore more favourable in the charged 
current case, and  the $\langle\sin (n\phi)\rangle$  are also
larger. Despite the significantly smaller cross section, a measurement of 
$T$-odd asymmetries in charged current 
DIS  might therefore be competitive to the measurement in the neutral
current case.

In summary, we have investigated the numerical magnitude of 
various asymmetries
in the angular distribution of hadrons in the final state of 
deep inelastic scattering, as determined by parton model
expressions. The resulting estimates for neutral current deep inelastic
scattering show that the 
$\langle \cos(n\phi)\rangle(P_T)$ asymmetries 
are typically of the order of a few per cent,
and should thus be easily measurable. The time-reversal-odd 
asymmetry $\langle \sin\phi\rangle(P_T)$ does hardly exceed $10^{-3}$ 
in neutral current interactions and $10^{-2}$ in charged current 
processes,
an experimental determination of it is therefore a challenging task. If a
substantially larger  $\langle \sin\phi\rangle(P_T)$ should be 
observed at HERA, it would be a clear indication for large
non-perturbative $T$-odd effects, as suggested in the
literature~\cite{nplit}.

\section*{Acknowledgements}
\noindent
The work of M.A.~was supported by DAAD. The authors would like to thank
G.~Kramer for several discussions throughout the project. 

\goodbreak

\newpage

\newpage

\newpage

\hfill

\end{document}